\documentstyle[12pt]{article}
\begin{document}
\begin{flushright} IM SB RAS NNA 7-96 \end{flushright}
\vspace{1cm}
\begin{center}
{\large MECHANISMS OF THE REACTION $\pi^-p\rightarrow a^0_0(980)n\rightarrow
\pi^0\eta n$ \\ AT HIGH ENERGIES} \\[2cm]
{\large N.N. Achasov and G.N. Shestakov}\\[0.7cm]
Laboratory of Theoretical Physics,\\
S.L. Sobolev Institute for Mathematics,\\
630090, Novosibirsk 90, Russia\\[3.5cm]
Abstract\\[0.7cm] \end{center}

The main dynamical mechanisms of the reaction $\pi^-p\rightarrow a^0_0
(980)n\rightarrow\pi^0\eta n$ at high energies, currently investigated at
Serpukhov and Brookhaven, are considered in detail. It is shown that the
observed forward peak in its differential cross section can be explained within
the framework of the Regge pole model only by the conspiring $\rho_2$ Regge
pole exchange. The tentative estimates of the absolute $\pi^-p\rightarrow a^0_0
(980)n\rightarrow\pi^0\eta n$ reaction cross section at $P_{lab}^{\pi^-}=18$
GeV/c are obtained: $\sigma\approx200$ nb and, in the forward direction,
$d\sigma/dt\approx940$ nb/GeV$^2$. The contribution of the one pion exchange,
which is forbidden by $G$-parity and which can rise owing to the $f^0_0(980)-
a^0_0(980)$ mixing, is also estimate. A role of the Regge cuts in the non-flip
helicity amplitude is briefly examined and a conclusion is made that the
contributions of the cuts have to be inessential in comparison with the
conspiring $\rho_2$ Regge pole exchange.
\\[0.7cm]
PACS number(s): 13.85.-t, 11.80.Cr, 12.40.Gg
\newpage
\section{Introduction}

In the $q\bar q$-model ($q$ is a light quark), every rotational excitation with
the orbital angular moment $L$ consists of four nonent: states $^{2S+1}L_J
=$ $^3L_{L-1}$, $^3L_L$, and $^3L_{L+1}$ with charge-parity $C=(-1)^{L+1}$ and
state $^1L_L$ with $C=(-1)^L$. However,
so far there is a white spot in the lower-lying
family with $L=2$ [1]. The non-strange members of the $^3D_2$ nonet with the
quantum numbers $I^G(J^{PC})=1^+(2^{--})$ and $0^-(2^{--})$, i.e. the $\rho_2$,
$\omega_2$, and $\phi_2$ mesons (the masses of which are expected near 1.7 and
1.9 GeV [2-4]), are not yet identified as peaks in corresponding multi-body
mass spectra [1]. The discussions of the possible reasons of this unusual
situation
are contained, for example, in Refs. [3,4]. However, the mass distributions are
not unique keepers of the information on the resonances. The resonance spectrum
is also reflected in the Regge behavior of the reaction cross sections at high
energies. At present the detailed investigations of the reaction
$\pi^-p\rightarrow\pi^0\eta n$ at $P^{\pi^-}_{lab}\approx40$ and 18 GeV/c are
carried out respectively at Serpukhov [5,6] and Brookhaven [7]. The $\pi^0\eta$
mass spectrum in this reaction is dominated by the $a^0_0(980)$ and $a^0_2(1320
)$ mesons [5-7]. In this connection, we should like to draw a special attention
to the reaction $\pi^-p\rightarrow a^0_0(980)n\rightarrow\pi^0\eta n$ because
its differential cross section near the forward direction can be dominated by
the Regge pole exchange with the quantum numbers of the ``lost'' $\rho_2$
meson. In general, this reaction is unique in that it involves only unnatural
parity exchanges in the $t$-channel.

The purpose of this paper is to describe in detail the main dynamical
mechanisms of the reaction $\pi^-p\rightarrow a^0_0(980)n\rightarrow\pi^0\eta
n$ in the Regge region. The paper is organized as follows. In Sec. II, we
present the reggeization of the $s$-channel helicity amplitudes of the reaction
$\pi^-p\rightarrow a^0_0(980)n$ and show that, in the framework of the Regge
pole model, the observed forward peak in its cross section [7] can be explained
by a very interesting and fine phenomenon such as conspiracy between the $\rho_2
$ trajectory and its daughter one. For the first time, the $\rho_2$ Regge
trajectory was introduced (at that time it was named $Z$) for the explanation
of the absence of a dip near the forward direction in $\rho^H_{00}d\sigma/dt(
\pi^-p\rightarrow\omega n)$ [8-10]. However, the nontrivial reason why the
$\rho_2$ Regge pole contribution in the $s$-channel amplitudes without helicity
flip in the nucleon vertex and with zero helicity of the $\omega$ meson do not
vanish at $t=0$, i.e. conspiracy of the Regge poles in $\pi^-p\rightarrow
\omega n$, did not discuss at all in Refs. [8-10]. Notice that, for the
similar cases, the necessary type of conspiracy was known [11-14] well before
the works [8-10]. Here we make up this omission by the example of the reaction
$\pi^-p\rightarrow a^0_0(980)n$. We present also the tentative estimate of the
$\pi^-p\rightarrow a^0_0(980)n\rightarrow\pi^0\eta n$ reaction cross section \
\footnote{Often the normalization of the reaction events turns out to be a
complicated problem. Probably in this connection, the experimental information
on the absolute cross section of the reaction $\pi^-p\rightarrow a^0_0(980)n
\rightarrow\pi^0\eta n$ is so far absent.} at
$P_{lab}^{\pi^-}=18$ GeV/c: $\sigma\approx200$ nb and in the forward direction
$d\sigma/dt\approx940$ nb/GeV$^2$. In Sec. III, we remind one more interesting
feature of the reaction $\pi^-p\rightarrow a^0_0(980)n\rightarrow\pi^0\eta n$
associated with the $f_0^0(980)-a_0^0(980)$ mixing [15] and estimate the
contribution of the one pion exchange which is possible owing to this mixing.
In Sec. IV, the role of the Regge cuts in the non-flip helicity amplitude is
briefly discussed. In Appendix, the conspiracy phenomenon is explained by the
example of the elementary $\rho_2$ exchange.

\section{Reaction $\pi^-p\rightarrow a^0_0(980)n$ at high energies in the
Regge pole model}

The $s$-channel helicity amplitudes of this reaction can be written as:
\begin{equation}
M_{\lambda_n\lambda_p}=\bar u_{\lambda_n}(p_2)\gamma_5\left[
-A-\frac{1}{2}\gamma^\mu(q_1+q_2)_\mu B\right]u_{\lambda_p}(p_1)\ ,
\end{equation}
where $q_1$, $p_1$, $q_2$, and $p_2$ are four-momenta of $\pi^-$, $p$,
$a^0_0$, and $n$ respectively, $\lambda_p$ and $\lambda_n$ are the proton and
neutron helicities, $A$ and $B$ are the invariant amplitudes depending on
$s=(p_1+q_1)^2$ and $t=(q_1-q_2)^2$ and free of kinematical singularities [16].
Using normalization $\bar uu=2m_N$ and taking the proton and the neutron as
``second particles'' [17] we obtain that, in the c.m. system,
\begin{equation}
M_{++}=-M_{--}=\cos(\theta/2)\left[A\sqrt{-t_{min}}-B\sqrt{-t_{max}s}\right]\ ,
\end{equation}
\begin{equation}
M_{+-}=+M_{-+}=\sin(\theta/2)\left[A\sqrt{-t_{max}}-B\sqrt{-t_{min}s}\right]\ ,
\end{equation}
where  $\theta$ is scattering angle, $t_{min}$ and $t_{max}$ are the values of
the variable $t$ at $\theta=0^\circ$ and $180^\circ$ respectively, $\sin(
\theta/2)=[-(t-t_{min})/4|\vec q_1||\vec q_2|]^{1/2}$, and
\begin{equation}
\frac{d\sigma}{dt}=\frac{1}{64\pi s|\vec q_1|^2}\left(|M_{++}|^2+|M_{--}|^2
\right)\ . \end{equation}
Eqs. (2) and (3) have the most simple form at high energies. Taking into
account that $A$ and $B$ at fixed $t$ and $s\gg m^2_N$ behave like $s^\alpha$
and $s^{\alpha-1}$ respectively (see below) and also $t_{min}\approx-m^2_N(m^
2_{a_0}-m^2_\pi)^2/s^2$ and $t_{max}\approx-s$, we get
\begin{equation} M_{++}\approx-sB\ , \ \ \ M_{+-}\approx\sqrt{-(t-t_{min})}A\ .
\end{equation}

The helicity amplitudes in the $t$-channel c.m. system $F_{\lambda_{\bar p}
\lambda_n}$ corresponding to the reaction $\pi^-a^0_0\rightarrow\bar pn$ have
the form \begin{equation}
F_{++}=-F_{--}=\sqrt{t}A+\frac{m_N(m^2_{a_0}-m^2_\pi)}{\sqrt{t}}B\ ,
\end{equation} \begin{equation}
F_{+-}=+F_{-+}=2|\vec q_t||\vec p_t|\sin\theta_tB\equiv\sqrt{\frac{\Phi(s,t,u)}
{t}}B\ , \end{equation}
In these equations, the $a^0_0(980)$ meson and neutron are taken as ``second
particles'', $\theta_t$ is scattering angle, $|\vec q_t|$ and $|\vec p_t|$
are the absolute values of the momenta of the particles in the initial and
final states respectively.
$\cos\theta_t=(s-u)/4|\vec q_t||\vec p_t|$, $u=2m^2_N+m^2_{a_0}+m^2_
\pi-s-t$; the equation $\Phi(s,t,u)=0$ gives the boundary of the physical
region. It is obvious, from Eqs. (6) and (7), that the helicity amplitudes
\begin{equation} G_{++}=\sqrt{t}F_{++} \qquad \mbox{and} \qquad
G_{+-}=\left(\sqrt{\frac{\Phi(s,t,u)}{t}}\right)^{-1}F_{+-}=B \end{equation}
are free of kinematical singularities. Their reggeization can be performed by
the usual way [18,19].

Constructing  the helicity amplitudes with  definite parity [18]
\begin{equation} G_{\lambda_{\bar p}\lambda_n}^{(\pm)}=G_{\lambda_{\bar p}
\lambda_n}\left[1\pm\eta_\pi\eta_{a_0}(-1)^{\lambda_{\bar p}-\lambda_n+|
\lambda_{\bar p}-\lambda_n|}\right]/2 \ , \end{equation}
we obtain that, because of the difference of the intrinsic parities of the
$\pi$ and $a_0$ mesons, $\eta_\pi$ and $\eta_{a_0}$, the amplitudes $G_{++}^{(
+)}$ and $G_{+-}^{(+)}$ identically vanish and thus $G_{++}\equiv G_{++}^{(-)}
$ and $G_{+-}\equiv G_{+-}^{(-)}$. Consequently, both amplitudes $G_{++}$ and
$G_{+-}$ have unnatural parity as it must be since each state of the $\pi a_0$
system with angular moment $J$ has parity $P_{\pi a_0}=(-1)^{J+1}$. It follows
from the parity conservation condition $P_{\bar pn}=(-1)^{L+1}=P_{\pi a_0}$,
where $L$ is angular moment of the $\bar pn$ system, that $L=J$ both for the
singlet $\bar pn$ spin state and for the triplet one. The amplitudes $G_{+-}$
and $G_{++}$ correspond to the triplet and singlet (because $G_{++}=-G_{--}$)
$\bar pn$ configurations
respectively. The $G$-parity conservation condition $(-1)^{L+S+I}=(-1)^{J+S+1}
=+1$, where $S$ and $I$ are spin and isospin of $\bar pn$, gives that in the
triplet (singlet) state only even (odd) values of $J$ are possible. The partial
wave expansions of $G_{++}$ and $G_{+-}$ are [18]:
\begin{equation} G_{++}=\sum_{J=1,3,...} (2J+1)f^J_{++}P_J(\cos\theta_t)\ ,
\qquad  G_{+-}=\sum_{J=2,4,...} (2J+1)f^J_{+-}\frac{P'_J(\cos\theta_t)}{\sqrt
{J(J+1)}}\ . \end{equation} Thus, the amplitude $G_{++}$ has to contain the
Regge pole exchanges with $I=1$, $G=+1$, signature $\tau=-1$, and
``naturality''
$\tau P=-1$. The high-lying Regge trajectory with such quantum numbers is the
$b_1$ trajectory (the well-known $b_1(1235)$ meson is its lower-lying
representative). The second independent amplitude $G_{+-}$ has to contain the
Regge pole exchanges with $I=1$, $G=+1$, $\tau=+1$, and $\tau P=-1$ and here
the
$\rho_2$ Regge trajectory is a leading one. Taking into account Eqs (10), the
contributions of the $b_1$ and $\rho_2$ Regge pole exchanges in the physical
region of the $s$-channel can be written as
\begin{equation}
G_{++}^{b_1}=\beta_{b_1}(t)\left(\frac{s}{s_0}\right)^{\alpha_{b_1}(t)}ie^{-i
\pi\alpha_{b_1}(t)/2}\ , \qquad
G_{+-}^{\rho_2}=\beta_{\rho_2}(t)\left(\frac{s}{s_0}\right)^{\alpha_{\rho_2}(t)
-1}e^{-i\pi\alpha_{\rho_2}(t)/2}\ , \end{equation}
where $\beta(t)$, $\alpha(t)$, and complex factors are residues, trajectories
and signatures of the corresponding Regge poles, and $s_0=1$ GeV$^2$. For
compensation of the nonphysical branch  points in $G_{+-}$ connected with
$1/\sqrt{J(J+1)}$ [see Eq. (10)], the factor $\sqrt{J(J+1)}$ has been extracted
from $f^J_{+-}$ [12].

Let us return to Eqs. (6) -- (8) and express the invariant amplitudes $A$ and
$B$ in terms of $G_{++}$ and $G_{+-}$. \begin{equation}
A=\frac{1}{t}\left[G_{++}-m_N\left(m^2_{a_0}-m^2_\pi\right)G_{+-}\right]\ ,
\end{equation} \begin{equation} B=G_{+-}\ . \end{equation}
To avoid the $1/t$ singularity in the invariant amplitude $A$ [see Eq. (12)],
it is necessary to complement the reggeization scheme by the conditions on
the behavior of the various contributions to $G_{\lambda_{\bar p}\lambda_n}$
as $t\rightarrow0$. Let us attempt to satisfy the analyticity of $A$ assuming
the $b_1$ and $\rho_2$ exchanges only and also that the amplitudes
$G^{b_1}_{++}$ and $G^{\rho_2}_{+-}$ do not vanish as $t\rightarrow0$. Then,
substituting Eq. (11) to (12) and going to the limit $t=0$, we obtain two
relations:  \begin{equation} \alpha_{b_1}(0)=\alpha_{\rho_2}(0)-1\ , \qquad
\beta_{b_1}(0)=m_N\left(m^2_{a_0}-m^2_\pi\right)\beta_{\rho_2}(0)\ ,
\end{equation} the first of which is rather silly because, at the usual
values of $\alpha_{b_1}(0)\approx-(0.05\div0.3)$ [8-10,20], it requires
$\alpha_{\rho_2}(0)\approx0.95\div0.7$ (also, for the linear $\rho_2$
trajecrory with the slope $\alpha'\approx0.8\div1$ GeV$^2$, it predicts the $
\rho_2$ mass $m_{\rho_2}\approx1.02\div1.27$ GeV).
For the $\rho_2$ trajectory heaving
unnatural parity, this is evidently ruled out. In fact, we conclude that there
is no way to make so that the residue of the $b_1$ exchange in Eq. (11) would
be finite at $t=0$. Of course, in order for the amplitude $A$ to be regular for
$t\rightarrow0$, one can accept that the amplitudes $G^{b_1}_{++}$ and
$G^{\rho_2}_{+-}$ are separately proportional to $t$.
\footnote{Using for $b_1\pi a_0$ and $b_1\bar NN$ interactions the effective
Lagrangians $L(b_1\pi a_0)\sim j^M_\mu b^\mu_1$ and $L(b_1\bar NN)\sim j^B_\mu
b^\mu_1$, where $j^M_\mu=(q_1-q_2)_\mu$ and $j^B_\mu=\bar u(p_2)\gamma_5(p_2-
p_1)_\mu v(p_1)$, one can easily verify that the contribution of the elementary
$b_1$ exchange to the amplitude $G_{++}$ for the reaction $\pi(q_1)+a_0(q_2)
\rightarrow\bar N(p_1)+N(p_2)$ turns out to be really proportional to $t$.} \
In this case, the amplitude $B$ in Eq. (13) and amplitude $M_{++}$ in Eq. (5)
caused by the $\rho_2$ exchange are also proportional to $t$. Then from Eqs.
(5) and (4), it follows immediately that, for $b_1$ and $\rho_2$ Regge pole
exchanges, $d\sigma/dt\sim |t|$ at small $|t|$. Thus this Regge pole model
predicts a dip near the forward direction in the $\pi^-p\rightarrow a^0_0n$
reaction cross section. On the contrary, the experiment [7] shows a clear
forward peak. This means that the amplitude $M_{++}$ with the quantum numbers
of the $\rho_2$ exchange in the $t$-channel does not vanish as $t\rightarrow0
$. In the framework of the Regge pole model, this can be attended only in the
case of a conspiracy of the $\rho_2$ Regge trajectory with its daughter one
($d$), which has to have the quantum numbers of the $b_1$ exchange. Let us
written down the contribution of such daughter trajectory near $t=0$ in the form
\begin{equation} G_{++}^d=\beta_d(t)\left(\frac{s}{s_0}\right)^{\alpha_d(t)}ie^
{-i\pi\alpha_d(t)/2}\ .\end{equation} Then, the amplitude $A$ should be regular
at $t=0$ [see Eq. (12)] if the following relations for the $\rho_2$, $d$ and
$b_1$ exchanges are valid:
\begin{equation} \alpha_d(0)=\alpha_{\rho_2}(0)-1\ , \qquad
\beta_d(0)=m_N\left(m^2_{a_0}-m^2_\pi\right)\beta_{\rho_2}(0)\ ,\ \footnote{
The detailed explanation of the conspiracy phenomenon by the example of the
elementary $\rho_2$ exchange is contained in Appendix.} \end{equation}
\begin{equation} \beta_{\rho_2}(0)\not=0 , \qquad \beta_{b_1}(t)\sim t\ .
\end{equation} Now neither the amplitude $B$ [see Eqs. (11), (13), and (17)]
nor the amplitude $M_{++}$ in (5) vanish at $t=0$. Moreover, asymptotically
(at large $s$) $M_{++}$ is dominated by the $\rho_2$ trajectory [(see Eqs. (5),
(11) and (13)] and $M_{+-}$ is dominated by the $b_1$ trajectory [(see Eqs.
(5), (11), (12), (15), and (16)]. As for the daughter trajectory contribution
and the non-asymptotic contribution of the $\rho_2$ trajectory
(which behaves as $\sim s^{\alpha_{\rho_2}(t)-1}$)
to the amplitude $A$ and consequently to $M_{+-}$ then they
can be neglected at all [see Eqs. (5), (11), (12), (15), and (16)]. Thus, on
the one hand, a role of the daughter trajectory, in practice, comes to only
the fact that the residue of the leading $\rho_2$ Regge pole $\beta_{\rho_2}(t)
$, owing to a conspiracy [see Eqs. (16) and (17)], does not vanish when $t
\rightarrow0$ and can be parametrized, for example, by the simplest exponential
form: $\beta_{\rho_2}(t)=-\gamma_{\rho_2}\exp(b^0_{\rho_2}t)/s_0$. At the same
time, the residue of the $b_1$ Regge pole in Eq. (11) has to be proportional to
$t$ [see Eq. (17)] and can be parametrized, for example, as: $\beta_{b_1}(t)=t
\gamma_{b_1}\exp(b^0_{b_1}t)/\sqrt{s_0}$. In our normalization, the constants
$\gamma_{\rho_2}$ and $\gamma_{b_1}$ are dimensionless. On the other hand, if
the daughter trajectory is parallel to the $\rho_2$ trajectory (as, for
example, in the Veneziano model) then, near 1.7 GeV, it should be expected a
state with the $b_1$ meson quantum numbers, which can be searched for in the
$a_0\pi$, $\omega\pi$, and $A_2\pi$ channels.

Thus, in the model with the $b_1$ and conspiring $\rho_2$ Regge poles, the
$s$-channel helicity amplitudes given by Eq. (5) can be written in the
following form convenient for fitting to the data: \begin{equation}
M_{++}=\gamma_{\rho_2}e^{b_{\rho_2}(s)t}\left(\frac{s}{s_0}\right)^{\alpha_
{\rho_2}(0)}e^{-i\pi\alpha_{\rho_2}(t)/2}\ , \end{equation} \begin{equation}
M_{+-}=\sqrt{-(t-t_{min})/s_0}\ \gamma_{b_1}e^{b_{b_1}(s)t}\left(\frac{s}{s_0}
\right)^{\alpha_{b_1}(0)}ie^{-i\pi\alpha_{b_1}(t)/2}\ ,  \end{equation}
where $\alpha_R(t)=\alpha_R(0)+\alpha'_Rt$, $b_R(s)=b^0_R+\alpha'_R\ln(s/
s_0)$, $R$ designates a Reggeon. Let us point out, as a guide, that $\alpha_{b_
1}(0)\approx-0.22$ and $\alpha_{\rho_2}(0)\approx-0.3$ for $\alpha'_{b_1}
\approx\alpha'_{\rho_2}\approx0.8$ GeV$^{-2}$, $m_{b_1}\approx1.235$ GeV, and
$m_{\rho_2}\approx1.7$ GeV. Using Eqs. (4), (18), and (19), we get
that, at large $s$, \begin{equation}
\frac{d\sigma}{dt}=\frac{1}{16\pi s^2}\left[\gamma^2_{\rho_2}
e^{2b_{\rho_2}(s)t}\left(\frac{s}{s_0}\right)^{2\alpha_
{\rho_2}(0)}+\left(\frac{t_{min}-t}{s_0}\right)\gamma^2_{b_1}
e^{2b_{b_1}(s)t}\left(\frac{s}{s_0}\right)^{2\alpha_{b_1}(0)}\right]\ .
\end{equation}

According to the Brookhaven data at $P^{\pi^-}_{lab}\approx18$
GeV/c [7] the $t$ distribution for events of the reaction $\pi^-p\rightarrow
a^0_0(980)n\rightarrow\pi^0\eta n$ is strongly peaked in the forward
direction (see Fig. 1). These data are fitted very well for $-t_{min}<-t<0.6$
GeV$^2$ by the single exponential form: \begin{equation}
dN/dt(\pi^-p\rightarrow a^0_0(980)n\rightarrow\pi^0\eta n)=C\ e^{\Lambda t}\ .
\end{equation}
The best fit (with $\chi^2\approx15.9$ for 22 degrees of freedom) is obtained
with $\Lambda=4.7$ GeV$^{-2}$ and $C=129$ events/GeV$^2$. It is shown in Fig.
1 by the solid curve.
Unfortunately, the Serpukhov data on $d\sigma/dt(\pi^-p\rightarrow a^0_0(980)n)
$ at 40 GeV/c are not yet presented. It is known only that, in the $\pi^0\eta$
invariant mass region $1\leq m_{\pi^0\eta}\leq1.2$ GeV and for $-t_{min}<-t<0.5
$ GeV$^2$ the differential cross section $d\sigma/dt(\pi^-p\rightarrow\pi^0\eta
n)$ has a similar peak in the forward direction [5]. Obviously, the Brookhaven
data are described formally by the single amplitude $M_{++}$ with the $\rho_
2$ exchange [see Eqs. (18) and (21)]. However, within $\pm(10-20)$\%
experimental uncertainties in $dN/dt$ [7], the form (21) can be effectively
reproduced for $-t_{min}<-t<0.6$ by means of Eq. (20) where the $b_1$
contribution should be also different from zero. The fit to the data [7] to
the form $dN/dt=C_1\exp(\Lambda_1t)+(t_{min}-t)C_2\exp(\Lambda_2t)$ with
$C_1=131$ events/GeV$^2$, $\Lambda_1=7.6$ GeV$^{-2}$, $C_2=340$ events/GeV$^2$,
and $\Lambda_2=5.8$ GeV$^{-2}$ gives a $\chi^2\approx15.9$ for 20 degrees of
freedom, and the corresponding curve is practically the same as the solid curve
in Fig. 1. The dashed and dotted curves in Fig. 1 show the $\rho_2$
[$C_1\exp(\Lambda_1t)$] and $b_1$ [$(t_{min}-t)C_2\exp(\Lambda_2t)$]
contributions separately, with the latter yields approximately 34\% of the
integrated cross section.
In order to determine rather accurately the parameters of
the simplest Regge pole model given by Eqs. (18) --  (20), the good data on
$d\sigma/dt(\pi^-p\rightarrow a^0_0(980)n)$ at several appreciably different
energies are needed. First of all, we have in mind the energies of the $\pi^-$
beams at Serpukhov ($\approx40$ GeV), Brookhaven ($\approx18$ GeV) and KEK ($
\approx10$ GeV). Notice that, according to the estimate $\sigma\sim(s)^{2
\alpha-2}$ with $\alpha\approx-0.3$, the $a^0_0(980)$ production cross section
at KEK should be approximately 36 times as large as one at Serpukhov.

So far the experimental information on the absolute values of the
$\pi^-p\rightarrow a^0_0(980)n\rightarrow\pi^0\eta n$ cross section is absent.
Nevertheless, in order to have an idea of this cross section, we shall estimate
$\sigma(\pi^-p\rightarrow a^0_0(980)n\rightarrow\pi^0\eta n)$ at $P^{\pi^-}_
{lab}=18$ GeV/c using the data on the reaction $\pi^-p\rightarrow a^0_2(1320)n$
and the Brookhaven data on the $\pi^0\eta$ mass spectrum in $\pi^
-p\rightarrow\pi^0\eta n$. According to Refs. [21,5] \begin{equation}
\sigma(\pi^-p\rightarrow a^0_2(1320)n)=18.5\pm3.7\ , \ 12.3\pm2.5\ , \ 2.7\pm
1.0\ \mbox{and}\ 0.395\pm0.080\ \ \mu\mbox{b} \end{equation}
at $P^{\pi^-}_{lab}=12$, 15, 40 and 100 GeV/c respectively. These data are
fitted quite well by the exponential function: \begin{equation}
\sigma(\pi^-p\rightarrow a^0_2(1320)n)\approx1.62\mbox{mb}[P^{\pi^-}_{lab}/(1
\mbox{GeV/c})]^{-1.8}\ . \end{equation}
Then at 18 GeV/c, $\sigma(\pi^-p\rightarrow a^0_2(1320)n\rightarrow\pi^0
\eta n)\approx1.29\ \mu b$ (here we have taken into account that $B(a_2(1320)
\rightarrow\pi\eta)\approx0.145$ [1]). Fig. 2 shows the Brookhaven data
(corrected by the registration efficiency) on the $\pi^0\eta$ mass spectrum in
the reaction $\pi^-p\rightarrow\pi^0\eta n$ at 18 GeV/c [7]. According to our
estimate the ratio $N(a_0(980))/N(a_2(1320))\approx1/6-1/7$, where $N(a_0
(980))$ and $N(a_2(1320))$ are the numbers of events, respectively, in the
$a_0(980)$ and $a_2(1320)$ peaks above background. Thus, one can expect that,
at $P^{\pi^-}_{lab}=18$ GeV/c,
$\sigma(\pi^-p\rightarrow a^0_0(980)n\rightarrow\pi^0\eta n)\approx 200$ nb
and also $[d\sigma/dt(\pi^-p\rightarrow a^0_0(980)n\rightarrow\pi^0\eta n)]_
{t\approx0}\approx940$ nb/GeV$^2$ according Eq. (21) with $\Lambda=4.7$ GeV$
^{-2}$. We emphasize that these estimates are rather tentative.

\section{One pion exchange in $\pi^-p\rightarrow a^0_0(980)n\rightarrow\pi^0
\eta n$}

It is now interesting to estimate the contribution to this reaction of the
reggeized one pion exchange (OPE), which is forbidden by $G$-parity. The
corresponding cross section has the form: \begin{equation}
\frac{d\sigma^{(OPE)}}{dtdm}=\frac{1}{\pi s^2}\ \frac{g^2_{\pi NN}}{4\pi}
\ \left[\frac{-te^{2b_\pi(s)(t-m^2_\pi)}}{(t-m^2_\pi)^2}\right]
m^3\rho_{\pi\pi}\sigma(\pi^+\pi^-\rightarrow\pi^0\eta)\ ,\end{equation} where
$g^2_{\pi NN}/4\pi\approx14.6$, $m$ is the invariant mass of the $\pi^0\eta$
system, $\rho_{\pi\pi}=(1-4m^2_\pi/m^2)^{1/2}$, $b_\pi(s)=b_\pi^0+\alpha'_\pi
\ln(s/s_0)$, $\alpha'_\pi\approx0.8$ GeV$^2$. This contribution arises owing
to the $f_0(980)-a^0_0(980)$ mixing violating isotopic invariance. The
$f_0(980)-a^0_0(980)$ mixing phenomenon and its possible manifestations in the
various reactions (for example, in $\pi^\pm N\rightarrow\pi^0\eta N$) were
considered in detail in the works [15]. Therefore, here we give only the
numerical estimates the absolute value of the OPE contribution at the
Brookhaven and Serpukhov energies.

Recall that the cross section of the reaction forbidden by $G$-parity $\pi^+\pi
^-\rightarrow\pi^0\eta$ [see Eq. (24)] is determined mainly by the transitions
$f_0(980)\rightarrow(K^+K^-+K^0\bar K^0)\rightarrow a^0_0(980)$ and, in the
region between the $K^+K^-$ and $K^0\bar K^0$ thresholds, which has a width of
8 MeV, it can be on the average from 0.4 to 1 mb [15]. Outside the region
$2m_{K^+}\leq m\leq2m_{K^0}$ $\sigma(\pi^+\pi^-\rightarrow\pi^0\eta)$ drops
sharply. The mentioned uncertainty in the estimate of $\sigma(\pi^+\pi^-
\rightarrow\pi^0\eta)$ reflects the spectrum of the model assumptions which
were made by many authors for the determination of the coupling constants of
the $f_0(980)$ and $a_0(980)$ resonances with the $\pi\pi$, $K\bar K$ and $\pi
\eta$ channels (see details in Refs. [15,22]). Note that the value of
$\sigma(\pi^+\pi^-\rightarrow\pi^0\eta)$ between the $K^+K^-$ and $K^0\bar K^0$
thresholds is controlled mainly by the production of ratios $(g^2_{f_0K^+K^-}/
g^2_{f_0\pi^+\pi^-})(g^2_{a^0_0K^+K^-}/g^2_{a_0\pi\eta})$ [15], where the
coupling constants $g$ determine the corresponding decay widths of the scalar
mesons, for example, $m\Gamma_{f_0\pi^+\pi^-}(m)=(g^2_{f_0\pi^+\pi^-}/16\pi)
\rho_{\pi\pi}$ and so on.

Taking these remarks into account and integrating Eq. (24) over $m$ from
$2m_{K^+}$ to $2m_{K^0}$, we get \begin{equation} \frac{d\sigma^{(OPE)}}{dt}
\approx(12-30)\mbox{nb}\ \left[\frac{-te^{\Lambda_\pi(t-m^2_\pi)}}{(t-m^2_\pi
)^2}\right] \end{equation} at $P^{\pi^-}_{lab}=18$ GeV/c and approximately five
times smaller at $P^{\pi^-}_{lab}=40$ GeV/c. For the reactions with the one
pion exchange, a typical slope in the considered energy region is: $\Lambda_\pi
(=2b_\pi(s))\approx(5-7)$ GeV$^{-2}$. Then, the integral of the function
confined in brackets in Eq. (25) over $t$ turns out to be approximately equal 1.
Hence we have $\sigma^{(OPE)}\approx(12-30)$ nb at 18 GeV/c. This is (6 --
15)\% of our estimate, $\sigma(\pi^-p\rightarrow a^0_0(980)n\rightarrow\pi^0
\eta n)\approx200$ nb, obtained at the end of Sec. II. Due to a smallness of
the $\pi$ meson mass, the $d\sigma^{(OPE)}/dt$ is enhanced for small $|t|$
[about (85 -- 90)\% of the integrated cross section $\sigma^{(OPE)}$ originate
from the region $0<-t<0.2$ GeV$^2$]. At the maximum situated near $t\approx-m^
2_\pi$, \begin{equation} (d\sigma^{(OPE)}/dt)_{t\approx-m^2_\pi}\approx(122-305
)\mbox{nb/GeV}^2\ .\end{equation} It can make up from 13 to 32.5\% of $[d\sigma
/dt(\pi^-p\rightarrow a^0_0(980)n\rightarrow\pi^0\eta n)]_{t\approx0}$ which
has been roughly estimated to be 940 nb/GeV$^2$ at 18 GeV/c (see the end of
Sec. II).

Thus, the violating $G$-parity OPE contribution is able to play a quite
appreciable role in the formation of the peak in $d\sigma/dt(\pi^-p\rightarrow
a^0_0(980)n\rightarrow\pi^0\eta n)$ near the forward direction. Note that
the features of the interference between the $\pi$ and $b_1$ exchanges in the
amplitude $M_{+-}$ were discussed in some detail in Ref. [15]. To
extract uniquely the amplitude $M_{+-}$ which can be dominated in the low $|t|$
range by the ``forbidden'' $\pi$ exchange, a polarized target and a measurement
of the neutron polarization in the reaction $\pi^-p\rightarrow a^0_0(980)n
\rightarrow\pi^0\eta n$ are necessary. It is also desirable to measure
the charge-symmetric reaction $\pi^-n\rightarrow a^0_0(980)p\rightarrow\pi^0
\eta p$ in which the $f^0_0(980)-a^0_0(980)$ interference has to have
opposite sign [15].

\section{Contributions of the Regge cuts}

The Regge cuts, just like the conspiring $\rho_2$ Regge pole, can give a
nonvanishing contributi- on to the amplitude $M_{++}$ for $t\rightarrow0$.
Generally speaking, it is difficult to distinguish the contributions of the
conspiring poles and cuts. However, the standard numerical estimates (such as
below) show that in the considered reaction the Regge cuts have to be
insignificant.

First of all, we carry out a classification of the two-Reggeon cuts
contributing to the amplitude $M_{++}$ of the reaction $\pi^-p\rightarrow a^0_0
(980)n$. According to Ref. [23], the signature of the cut is given by $\tau_c=
\tau_1\tau_2$, where $\tau_1$ and $\tau_2$ are the signatures of the Regge
poles associated with the cut. The signature of the amplitude $M_{++}$ is
positive and therefore the $\tau_1$ and $\tau_2$ must be equal. Then, it is
found that the two-Reggeon cuts associated with the Regge poles having the
equal and opposite ``naturalities'' $(\tau P)$ have a principle different
behavior
as $t\rightarrow0$. Parity conservation gives that the cuts with $(\tau_1P_1)
(\tau_2P_2)=-1$ do not vanish as $t\rightarrow0$ [24]. Among these are the
$a_2\pi$, $\rho b_1$, and $\omega a_1$ cuts and also the P$\rho_2$ cut, where
P is the Pomeron. The cuts with $(\tau_1P_1)(\tau_2P_2)=+1$ give vanishing
contributions to $M_{++}$ as $t\rightarrow0$ (they turn out to be proportional
to $t$) [24]. In this group, the $\rho\rho$ and $a_2a_2$ cuts are leading at
large $s$.

The amplitude of the two-Reggeon cut associated with the $R_1$ and $R_2$ Regge
pole exchanges can be calculated in the absorption model approximation by the
formula [25-27]: \begin{equation} M^{R_1R_2}_{ab\rightarrow cd}(s,t)=\frac{i}
{8\pi^2s}\int d^2k_\bot\sum_{e,f}M^{R_1}_{ab\rightarrow ef}(s,\vec q-\vec k_
\bot)\ M^{R_2}_{ef\rightarrow cd}(s,\vec k_\bot)\ ,\end{equation} that is,
considering the $R_1R_2$ cut contribution as a process of a double quasielastic
rescattering. In Eq. (27), $\vec k_\bot$ and $\vec q$ are the momenta
transferred from the particle $e$ to the particle $c$ and from $a$ to $c$
respectively, $\vec q\,^2\approx-t$, the intermediate states $e$ and $f$
represent stable particles or narrow resonances. The accumulated wide
experience of the work with the two-Reggeon cuts shows that reasonable
estimates can be obtained considering the contributions of the simplest
(lowest-lying) intermediate states. The calculation methods of the two-Reggeon
cuts are well known (see, for example, Refs. [25-27,19,24]). Therefore,
omitting details, we go at once to the discussion of the final results. All
these are concerned with $P_{lab}^{\pi^-}=18$ GeV/c.

Begin with the $a_2\pi$ cut. Taking into account in Eq. (27) the low-lying
$\eta n$ intermediate state, we get the following contributions of the $a_2\pi$
cut to the $(d\sigma/dt)_{t\approx0}$ and integral cross section $\sigma$
of the reaction $\pi^-p\rightarrow a^0_0(980)n\rightarrow\pi^0\eta n$:
\begin{eqnarray} \left(\frac{d\sigma^{a_2\pi}}{dt}\right)_{t\approx0}=
\frac{I(m^2_\pi(\tilde b_{a_2}+\tilde b_\pi))}{4\pi|\tilde b_
{a_2}+\tilde b_\pi|^2}\left(\frac{1}{t}\frac{d\sigma^{a_2}_{hf}}{dt}\right)
_{t\approx0}\left(\frac{m^4_\pi}{t}\frac{d\sigma^\pi}{dt}\right)_{t\approx0}
\approx \nonumber \\ \approx25\ (\mbox{nb/GeV}^2)\ B(a_0^0(980)\rightarrow\pi
\eta)\ , \ \ \ \ \ \ \ \ \ \ \ \  \end{eqnarray} \begin{equation} \sigma^{a_2
\pi}\approx3.4\ (\mbox{nb})\ B(a_0^0(980)\rightarrow\pi\eta)\ . \end{equation}
Here $\tilde b_R=b_R-i\pi\alpha'_R/2$ (the argument $s$ of the slope $b_R$
is omitted from this moment), $d\sigma^{a_2}_{hf}/dt$ is the part of the $\pi^-
p\rightarrow\eta n$ differential cross section caused by the $a_2$ Regge pole
exchange with a helicity-flip in the nucleon vertex, $d\sigma^\pi/dt$  is the
differential cross section of the reaction $\eta n\rightarrow a^0_0(980)n
\rightarrow\pi^0\eta n$ caused by the $\pi$ Regge pole exchange. According the
Fermilab data on $\pi^-p\rightarrow\eta n$ [28], $[(1/t)d\sigma^{a_2}_{hf}/dt]_
{t\approx0}\approx555$ $\mu$b/GeV$^4$, $b_{a_2}\approx4.18$ GeV$^{-2}$, and
$\alpha'_{a_2}\approx0.8$ GeV$^{-2}$ ($\alpha_{a_2}(0)\approx0.371$). For the
reaction with the $\pi$ exchange, $[(m^4_\pi/t)d\sigma^\pi/dt]_{t\approx0}=g^2_
{\pi NN}(g^2_{a_0\pi\eta}/16\pi)\exp(-2b_\pi m^2_\pi)B(a^0_0(980)\rightarrow
\pi\eta)$, where $g^2_{a_0\pi\eta}/16\pi=\Gamma_{a_0\eta\pi}m_{a_0}/\rho_{\eta
\pi}$ and $\rho_{\eta\pi}=[(1-(m_\eta-m_\pi)^2/m_{a_0}^2)(1-(m_\eta+m_\pi)^2/m_
{a_0}^2)]^{1/2}$. According the Particle Data Group [22], the width $\Gamma_{a_
0\eta\pi}$ can be from 50 to 300 MeV.  We use its maximal value. Then,
$g_{a_0\eta\pi}^2/16\pi\approx0.454$ GeV$^2$. Also we assume that $\alpha'_\pi
\approx0.8$ GeV$^{-2}$ and  $b_\pi\approx3.5$ GeV$^{-2}$. The factor $I(m^2_\pi
(\tilde b_{a_2}+\tilde b_\pi))$ in Eq. (28) has the form $|1+z\exp(z)Ei(-z)|^2
$, where $z=m^2_\pi(\tilde b_{a_2}+\tilde b_\pi)$ and $Ei(-z)$ is the integral
exponential function. Here we have $I(m^2_\pi(\tilde b_{a_2}+\tilde b_\pi))
\approx0.55$.

Because $B(a^0_0(980)\rightarrow\pi\eta)<1$, then Eqs. (28) and (29) give,
respectively, less than 2.7\% and 1.7\% of the expected values
$[d\sigma/dt(\pi^-p\rightarrow a^0_0(980)n\rightarrow\pi^0\eta n)]_{t\approx0}
\approx940$ nb/GeV$^2$ and $\sigma(\pi^-p\rightarrow a^0_0(980)n\rightarrow
\pi^0\eta n)\approx200$ nb \ \footnote{Note that the $a_2\pi$ cut contribution
to $d\sigma/dt$ has a minimum around $t\approx-0.4$ GeV$^2$ and decreases by
approximately 54 times over the range of $t$ from 0 to $-0.4$ GeV$^2$. However,
experimentally $d\sigma/dt$ falls by a factor of 6.5 in this $t$-range and has
not the minimum [see Eq. (21)].}. Even though we magnify these numbers by an
order of magnitude (attributing the enhancement to the contributions of the
other intermediate states), all the same they would be appreciably smaller of
the expected values.

Turn to the $\rho b_1$ cut. The contribution of the low-lying $\pi^-p$
intermediate state is convenient- ly represented in the following form:
\begin{eqnarray} \left(\frac{d\sigma^{\rho b_1}}{dt}\right)_{t\approx0}=
\frac{1}{4\pi|\tilde b_\rho+\tilde b_{b_1}|^4}
\ \left(\frac{1}{t}\frac{d\sigma^\rho_{hf}}{dt}\right)_{t\approx0}\
\left(\frac{1}{t}\frac{d\sigma^{b_1}}{dt}\right)_{t\approx0}=\nonumber \\
=\frac{4}{\pi}\ \frac{b^2_\rho b^2_{b_1}}{|\tilde b_\rho+\tilde b_{b_1}|^4}
\ \sigma^\rho_{sf}\ \sigma^{b_1}<0.5\ \mbox{nb/GeV}^2 \ , \ \ \ \ \ \
\end{eqnarray} where $\sigma^\rho_{hf}$ is the $\pi^-p\rightarrow\pi^-p$
cross section with the proton helicity-flip caused by the $\rho$ Regge pole
exchange, $\sigma^{b_1}$ is the cross section of the reaction $\pi^-p
\rightarrow a^0_0(980)n\rightarrow\pi^0\eta n$ associated with the $b_1$
Regge pole exchange. The limitation (30) has been obtained in terms of the
following inequalities:
$\sigma^\rho_{hf}<\sigma(\pi^-p\rightarrow\pi^0n)/2\approx12.5$ $\mu$b [29],
$\sigma^{b_1}<\sigma(\pi^-p\rightarrow a^0_0(980)n\rightarrow\pi^0\eta n)
\approx200$ nb, $b^2_\rho b^2_{b_1}/|\tilde b_\rho+\tilde b_{b_1}|^4<1/16$.
Thus, the $\rho b_1$ cut contribution should be considered as a whole as very
small.

The $\omega a_1$ cut is more difficult to estimate because there appear the
amplitudes with the $a_1$ Regge pole exchange which are directly unobservable
by experiment. Consider the contributions of two simplest intermediate states
$\rho^-p$ and $b^-_1p$. At the expense of the $b^-_1p$ intermediate state, we
have \begin{eqnarray}
\left(\frac{d\sigma^{\omega a_1}}{dt}\right)_{t\approx0}\approx
\frac{1}{4\pi|\tilde b_\omega+\tilde b_{a_1}|^2}
\ \left(\frac{d\sigma^\omega}{dt}\right)_{t\approx0}\
\left(\frac{d\sigma^{a_1}}{dt}\right)_{t\approx0}\approx
\frac{1}{\pi}\ \frac{b_\omega b_{a_1}}{|\tilde b_\omega+\tilde b_{a_1}|^2}
\ \sigma^\omega\ \sigma^{a_1}< \nonumber \\
<\frac{1}{4\pi}\ \sigma^\omega\ \sigma^{a_1}
\approx10\ (\mbox{nb/GeV}^2)\ B(a_0^0(980)\rightarrow\pi\eta) \ ,
\ \ \ \ \ \ \ \ \ \ \ \ \ \ \ \ \ \ \ \ \ \       \end{eqnarray}
where $\sigma^\omega$ and $\sigma^{a_1}$ are cross sections of the reactions
$\pi^-p\rightarrow b^-_1p$ and $b^-_1p\rightarrow a^0_0(980)n\rightarrow\pi^0
\eta n$ with the $\omega$ and $a_1$ Regge pole exchanges respectively (there
are not helicity-flip in the nucleon vertices and the intermediate $b^-_1$
meson has in the main helicity zero [30]). $\sigma^\omega\approx[\sigma(\pi^+p
\rightarrow b^+_1p)+\sigma(\pi^-p\rightarrow b^-_1p)-\sigma(\pi^-p\rightarrow
b^0_1n)]/2\approx20$ $\mu$b [29-31]. To estimate $\sigma^{a_1}$ one can
virtually be guided by only the data on the reaction $\pi^-p\rightarrow\rho^0n$
at 17.2 GeV/c [32]. The exchanges with the $a_1$ quantum numbers make up
approximately 4\% of this reaction cross section ($\approx20\%$ in the
amplitude), i.e., $\approx2.5$ $\mu$b [33]. To obtain the estimate (31), we
have use a rough assumption: $\sigma^{a_1}(b^-_1p\rightarrow a^0_0n)\approx
\sigma^{a_1}(\pi^-p\rightarrow\rho^0n)$. Similarly, one can obtain for the
contribution of the $\rho^-p$ intermediate state with the transverse polarized
$\rho^-$ meson that $(d\sigma^{\omega a_1}/dt)_{t\approx0}<(\sigma^\omega\sigma
^{a_1}/4\pi)\approx7.5$ (nb/GeV$^2)B(a_0^0(980)\rightarrow\pi\eta)$, where
$\sigma^\omega\approx[\sigma(\pi^+p\rightarrow\rho^+p)
+\sigma(\pi^-p\rightarrow\rho^-p)-\sigma(\pi^-p\rightarrow\rho^0n)]/2\approx
15$ $\mu$b [29,34] and, for the $\rho^-p\rightarrow a^0_0(980)n$ reaction cross
section with the $a_1$ exchange, we take simply the same 2.5 $\mu$b as just
above. A relative sign of the $\rho^-p$ and $b^-_1p$ intermediate state
contributions is unknown. As a result for the $\omega a_1$ cut, we have a very
rough limitation: \begin{equation}
(d\sigma^{\omega a_1}/dt)_{t\approx0}<35\ (\mbox{nb/GeV}^2)\ B(a_0^0(980)
\rightarrow\pi\eta). \end{equation}

As mentioned above, the contributions of the $\rho\rho$ and $a_2a_2$ cuts to
$M_{++}$ vanish as $t\rightarrow0$. However, the absorption corrections to
these contributions, i.e., the $\rho\rho$P and $a_2a_2$P cuts, are finite as
$t\rightarrow0$  \footnote{Formulae for the such type cuts were obtained in
Ref. [24].}. The estimates of the $\rho\rho$P and $a_2a_2$P cut contributions
to $[d\sigma/dt(\pi^-p\rightarrow a^0_0(980)n\rightarrow\pi^0\eta n)]_{t\approx
0}$, quite similar done above, show that each of ones does not exceed by itself
2 nb/GeV$^2$, that is, very small. Moreover, a strong compensation between the
$\rho\rho$ and $a_2a_2$ cuts (and analogously between $\rho\rho$P and $a_2a_2$P
cuts) takes place within the framework of the $\rho-a_2$ exchange degeneracy
hypothesis because the productions of the Regge pole signature factors for
these cuts are opposite in sign.

As for the P$\rho_2$ cut, the absorption correction of this type accompany any
Regge pole exchange. In the cases that the pole contributions do not vanish as
$t\rightarrow0$ (beyond general kinematic and factorization requirements),
these corrections are effectively unimportant at least for the description of
the differential cross sections. The reactions $\pi^-p\rightarrow\pi^0n$ and
$\pi^-p\rightarrow\eta n$ give classical examples of this situation. At small
$|t|$ and in a wide energy region, the differential cross sections of these
reactions are described remarkably well by a simple Regge pole model with the
linear $\rho$ and $a_2$ trajectories [28,35].

\section{Conclusion}

We have considered the main dynamical mechanisms of the reaction $\pi^-p
\rightarrow a^0_0(980)n\rightarrow\pi^0\eta n$ at high energies and shown that
the observed peak in its differential cross section in the forward direction
can be explained within the framework of the Regge pole model only by a
conspiracy of the $\rho_2$ trajectory with its daughter one. Notice that there
realizes another type of conspiracy in the well known cases of the reactions
$\gamma p\rightarrow\pi^+n$ and $pn\rightarrow np$ (in which the corresponding
peaks in the forward direction are usually described in terms of the Regge
cuts [19,25,27]) than in the reaction $\pi^-p\rightarrow a^0_0(980)n$ [12].
We have also obtained the estimates of the $\pi^-p\rightarrow a^0_0(980)n
\rightarrow\pi^0\eta n$ reaction cross section at $P_{lab}^{\pi^-}=18$ GeV/c
and of the OPE contribution which can be caused by the $f^0_0(980)-a^0_0(980)$
mixing. Examining the Regge cut contributions to the non-flip helicity
amplitude, we have come to conclusion that they have to be inessential in
comparison with the conspiring $\rho_2$ Regge pole exchange.

Certainly, it would be very interesting to find some signs of the $\rho_2$
state and its daughter state with the $b_1$ meson quantum numbers, for example,
in the $a_0\pi$, $\omega\pi$, and $A_2\pi$ mass spectra around 1.7 GeV in the
reactions induced by $\pi^\mp$ mesons or in $\bar NN$ annihilation.
\vspace{0.3cm}
\begin{flushleft} {\large\bf{Acknowledgements}} \end{flushleft}

N.N. Achasov thanks A.R. Dzierba for many discussions.

This work was partly supported by the Russian Foundation for
Fundamental Research- es Grant No. 94-02-05 188, the Grant No. 96-02-00548,
and the INTAS Grant No. 94-3986.


\vspace{0.6cm}
\begin{flushleft} {\large\bf{Appendix}} \end{flushleft}

Let us explain a conspiracy phenomena by example of the elementary $\rho_2$
exchange in the reaction $\pi(q_1)+a_0(q_2)\rightarrow\rho_2(q)\rightarrow\bar
N(p_1)+N(p_2)$. The effective Lagrangians for the $\rho_2\pi a_0$ and $\rho_2
\bar NN$ interactions can be written as (we omit the coupling constants)
\begin{equation} L(\rho_2\pi a_0)=j^M_{\mu\nu}\rho_2^{\mu\nu}\ , \qquad
j^M_{\mu\nu}=Q_\mu Q_\nu \end{equation} \begin{equation}
L(\rho_2\bar NN)=j^B_{\mu\nu}\rho_2^{\mu\nu}\ , \qquad
j^B_{\mu\nu}=\bar u(p_2)\gamma_5\frac{1}{4}(\gamma_\mu P_\nu+\gamma_\nu P_\mu)
v(p_1)\ , \end{equation}
where $P=p_2-p_1$, $Q=q_1-q_2$, $q=q_1+q_2=p_1+p_2$ \ ($PQ=s-u$, $Pq=0$, $Qq=
m^2_\pi-m^2_{a_0}$, $q^2=t$). The helicity amplitudes of the process $\pi^-
a^0_0\rightarrow\rho^-_2\rightarrow\bar pn$ are then
\begin{equation} F_{\lambda_{\bar p}\lambda_n}=V^{\mu\nu}_{\lambda_{\bar p}
\lambda_n}\ \frac{\Pi_{\mu\nu\mu'\nu'}}{q^2-m^2_{\rho_2}}\ Q^{\mu'}Q^{\nu'}\ ,
\end{equation} where
\begin{equation} V^{\mu\nu}_{\lambda{\bar p}\lambda_n}=
\bar u_{\lambda_n}(p_2)\gamma_5\frac{1}{4}(\gamma^\mu P^\nu+\gamma^\nu P^\mu)
v_{\lambda_{\bar p}}(p_1)\ , \end{equation}
\begin{equation} \Pi_{\mu\nu\mu'\nu'}=\frac{1}{2}\pi_{\mu\mu'}\pi_{\nu\nu'}+
\frac{1}{2}\pi_{\mu\nu'}\pi_{\nu\mu'}-\frac{1}{3}\pi_{\mu\nu}\pi_{\mu'\nu'}\ ,
\qquad \pi_{\mu\nu}=g_{\mu\nu}-q_\mu q_\nu/m^2_{\rho_2}\ . \end{equation}
Off mass shall ($q^2\not=m^2_{\rho_2}$), the tensor $\Pi_{\mu\nu\mu'\nu'}$
is not the spin-2 projection operator but contains the contributions of the
lower (daughter) spins. In the case of coupling to nonconserved tensor
currents, these daughter contributions appear in the physical amplitudes. In
the case under consideration, the elementary $\rho_2$ exchange with spin 2
is accompanied by the spin-1 contribution only (the spin-0 contribution is
uncoupled to the $\bar NN$ system because it has the exotic quantum numbers
$I^G(J^{PC})=1^+(0^{--})$). Using the relation $j^B_{\mu\nu}g^{\mu\nu}=0$ and
$j^B_{\mu\nu}q^\mu q^\nu=0$, Eq. (35) can be rewritten in the form
\begin{equation} F_{\lambda_{\bar p}\lambda_n}=V^{\mu\nu}_{\lambda_{\bar p}
\lambda_n}\ \frac{P^{(2)}_{\mu\nu\mu'\nu'}+[(q^2-m^2_{\rho_2})/m^2_{\rho_2}]
P^{(1)}_{\mu\nu\mu'\nu'}}{q^2-m^2_{\rho_2}}\ Q^{\mu'}Q^{\nu'}\ ,
\end{equation} where the tensors \begin{equation}
P^{(2)}_{\mu\nu\mu'\nu'}=\frac{1}{2}\theta_{\mu\mu'}\theta_{\nu\nu'}+
\frac{1}{2}\theta_{\mu\nu'}\theta_{\nu\mu'}-\frac{1}{3}\theta_{\mu\nu}\theta_
{\mu'\nu'}\  \qquad (\theta_{\mu\nu}=g_{\mu\nu}-q_\mu q_\nu/q^2)\ ,
\end{equation} and \begin{equation}
P^{(1)}_{\mu\nu\mu'\nu'}=\frac{1}{2q^2}\left[\theta_{\mu\mu'}q_\nu q_{\nu'}+
\theta_{\nu\nu'}q_\mu q_{\mu'}+\theta_{\mu\nu'}q_\nu q_{\mu'}+\theta_{\nu\mu'}
q_\mu q_{\nu'}\right] \end{equation} are the spin-2 and spin-1 projection
operators respectively [36] [$(P^{(2)})^2=P^{(2)}$,\  $(P^{(1)})^2=P^{(1)}$,\
$P^{(2)\mu\nu}_{\ \mu\nu}=5$,\  $P^{(1)\mu\nu}_{\ \mu\nu}=3$]. The spin-2 and
spin-1 parts of the elementary $\rho_2$ exchanges give the contributions to
the amplitudes $F_{+-}$ and $F_{++}$ respectively. This immediately follows
from an explicit form of the angular and threshold behaviors of these
amplitudes. \begin{equation}
F_{+-}=-8|\vec q_t|^2|\vec p_t|^2\cos\theta_t\sin\theta_t\ \frac{1}{t-m^2
_{\rho_2}}\equiv-\sqrt{\frac{\Phi(s,t,u)}{t}}\ \frac{s-u}{t-m^2_{\rho_2}}\ ,
\end{equation} \begin{equation}
F_{++}=4|\vec q_t||\vec p_t|\cos\theta_t\ \frac{m_N(m^2_{a_0}-m^2_\pi)}
{\sqrt{t}m^2_{\rho_2}}\equiv\frac{(s-u)m_N(m^2_{a_0}-m^2_\pi)}{\sqrt{t}m^2_
{\rho_2}}\ . \end{equation}
Both amplitudes are singular as $t\rightarrow0$ as $1/\sqrt{t}$. Now we go
from the amplitudes $F_{+-}$ and $F_{++}$ given by Eqs. (41) and (42) to the
amplitudes $G_{++}$ and $G_{+-}$ [see Eq. (8)]. Substituting $G_{++}$ and
$G_{+-}$ to Eq. (12), we see that, owing to the compensation between these
helicity amplitudes having different quantum numbers, the $1/t$ singularity in
the invariant amplitude $A$ is cancelled out:
\begin{eqnarray}
A=\frac{1}{t}\left[G_{++}-m_N\left(m^2_{a_0}-m^2_\pi\right)G_{+-}\right]=
\ \ \ \ \ \ \ \ \ \ \ \ \ \ \ \ \ \ \ \ \ \ \ \ \ \ \ \ \ \ \ \ \ \nonumber\\
=4|\vec q_t||\vec p_t|\cos\theta_t\ \frac{m_N(m^2_{a_0}-m^2_\pi)}
{t}\left(\frac{1}{m^2_{\rho_2}}+\frac{1}{t-m^2_{\rho_2}}\right)=\frac{s
-u}{t-m^2_{\rho_2}}\ \frac{m_N(m^2_{a_0}-m^2_\pi)}{m^2_{\rho_2}}\ .
\end{eqnarray}
It takes place automatically since the conspiracy condition, \begin{equation}
G_{++}=m_N(m^2_{a_0}-m^2_\pi)G_{+-} \qquad \mbox{at}\ \ t=0 \ , \end{equation}
for the elementary $\rho_2$ exchange is exactly fulfilled. The froward peak in
$d\sigma/dt$ is provided by the second invariant amplitude $B=G_{+-}=-(s-u)/(t
-m^2_{\rho_2})$, which, as seen, does not vanish at $t=0$ [see Eqs.
(4) and (5)]. As for the contribution of the amplitude $A$ to $d\sigma/dt$
then it is an $s$ times smaller at large $s$ and therefore it can be neglected
[see Eqs. (43), (5) and (4)].

\vspace{1cm}
\begin{center} \large{\bf Figure captions} \end{center} \vspace{0.5cm}

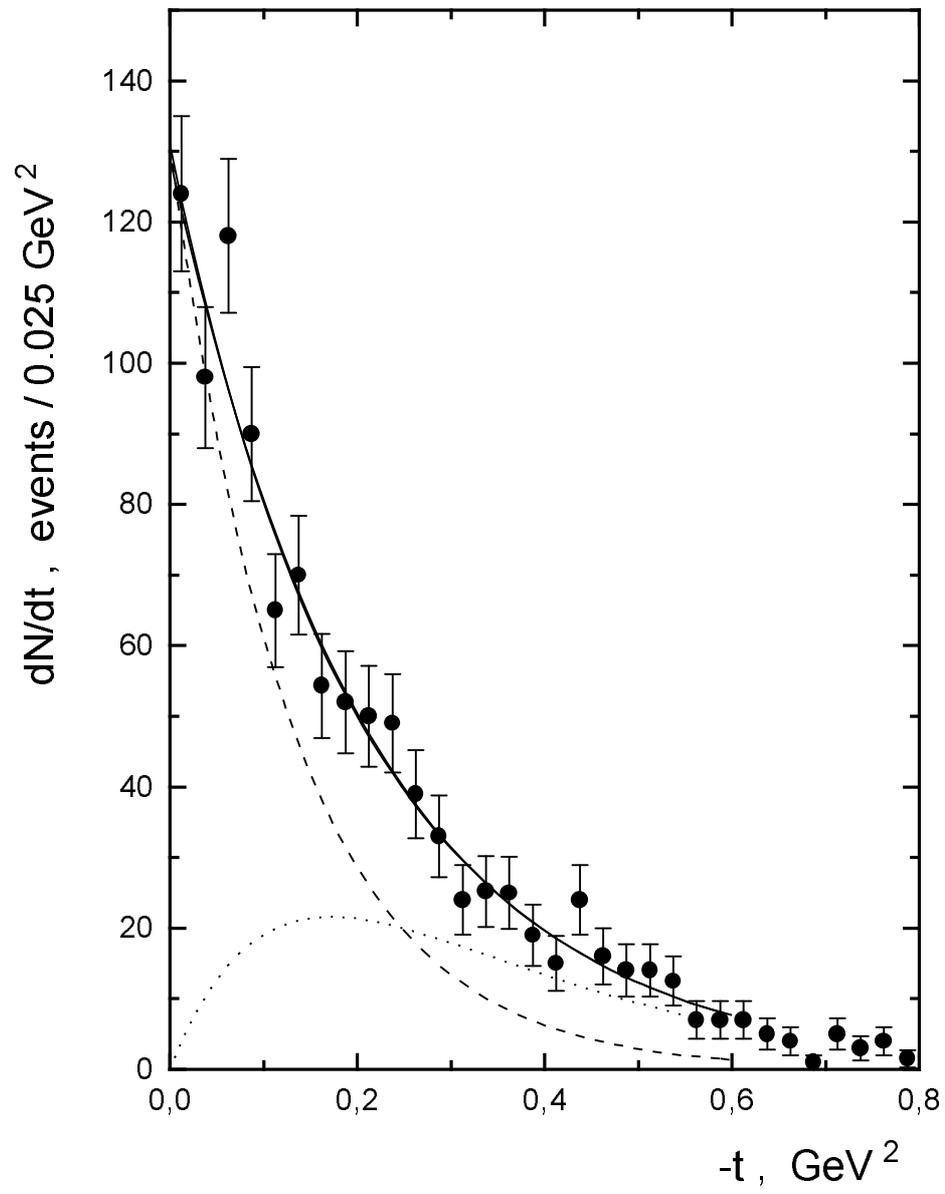
{\bf Fig. 1.} The $t$ distribution for the reaction $\pi^-p\rightarrow a^0_0(
980)n\rightarrow\pi^0\eta n\rightarrow4\gamma n$ at 18 Gev/c measured at
Brookhaven [7]. The fits are described in the text. \vspace{0.3cm}

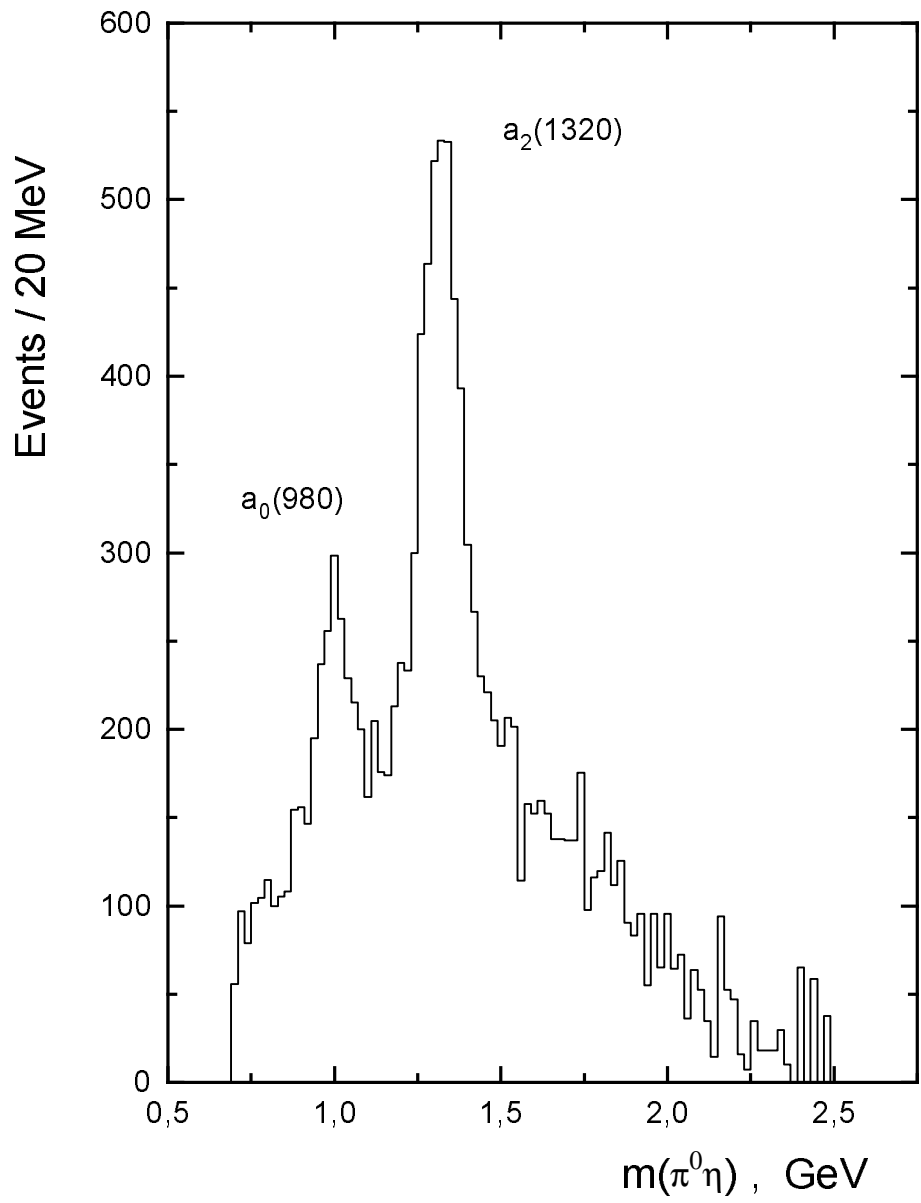
{\bf Fig. 2.} The $\pi^0\eta$ mass spectrum for the reaction $\pi^-p\rightarrow
\rightarrow\pi^0\eta n\rightarrow4\gamma n$ at 18 Gev/c measured at
Brookhaven [7].

\end{document}